\documentclass[
  journal=largetwo,
  manuscript=article-type,
  year=2025,
  volume=37,
]{cup-journal}

\usepackage{amsmath}
\usepackage{amssymb,microtype,siunitx,booktabs}
\usepackage{aas-macros}

\usepackage{mathrsfs}
\usepackage{hyperref}
\usepackage{orcidlink}

\hypersetup{
    colorlinks=true,
    linkcolor=blue,
    filecolor=blue,
    urlcolor=blue,
    citecolor=blue,
    pdfborder={0 0 0}
}

\title{On the redshift evolution of the spin parameter in cosmological simulations}

\author{Tomas Riera}
\affiliation{Departamento de F\'isica Te\'{o}rica, M\'{o}dulo 15, Facultad de Ciencias, Universidad Aut\'{o}noma de Madrid, 28049 Madrid, Spain}
\email[A. Knebe]{alexander.knebe@uam.es}

\author{Alexander Knebe\textsuperscript{\href{https://orcid.org/0000-0003-4066-8307}{\orcidlink{0000-0003-4066-8307}}}}
\affiliation{Departamento de F\'isica Te\'{o}rica, M\'{o}dulo 15, Facultad de Ciencias, Universidad Aut\'{o}noma de Madrid, 28049 Madrid, Spain}
\alsoaffiliation{Centro de Investigaci\'{o}n Avanzada en F\'isica Fundamental (CIAFF), Facultad de Ciencias, Universidad Aut\'{o}noma de Madrid, 28049 Madrid, Spain}
\alsoaffiliation{International Centre for Radio Astronomy Research, University of Western Australia, 35 Stirling Highway, Crawley, Western Australia 6009, Australia}

\author{Chris Power\textsuperscript{\href{https://orcid.org/0000-0002-4003-0904}{\orcidlink{0000-0002-4003-0904}}}}
\affiliation{International Centre for Radio Astronomy Research, University of Western Australia, 35 Stirling Highway, Crawley, Western Australia 6009, Australia}
\alsoaffiliation{Australian Research Council (ARC) Centre of Excellence for All Sky Astrophysics in 3 Dimensions (ASTRO 3D)}

\author{Robert Adriel Mostoghiu Paun\textsuperscript{\href{https://orcid.org/0000-0002-5425-0641}{\orcidlink{0000-0002-5425-0641}}}}
\affiliation{Centre for Astrophysics and Supercomputing, Swinburne University of Technology, PO Box 218, Hawthorn VIC 3122, Australia} 
\alsoaffiliation{Australian Research Council (ARC) Centre of Excellence for Dark Matter Particle Physics (CDMPP)}

\author{Adam Ussing\textsuperscript{\href{https://orcid.org/0009-0009-4592-6628}{\orcidlink{0009-0009-4592-6628}}}}
\affiliation{Centre for Astrophysics and Supercomputing, Swinburne University of Technology, PO Box 218, Hawthorn VIC 3122, Australia}


\begin{document}

\begin{abstract}
{Although the spin parameter of dark matter halos is well known to follow a log-normal distribution at fixed epoch, its quantitative redshift evolution - encompassing both the mean and the dispersion - remains only partially explored. Prior studies either lack the mass resolution required to establish reliable evolutionary trends or do not provide analytical relations that enable forward modelling. Using a suite of $\Lambda$CDM N-body simulations with controlled resolution across the redshift range $0\leq z \leq 5$, we characterise the evolution of the mean and dispersion of the Peebles ($\lambda$) and Bullock ($\lambda^\prime$) definitions of spin. We find a mild but statistically robust linear evolution for $\ln\lambda$ and a non‑monotonic trend with a turnover at $z\approx 1-2$ for $\ln\lambda^\prime$, which we verify are unaffected by mass resolution of choice of halo definition. We provide closed-form fitting functions for these trends that allow modellers to draw physically motivated spin values at any redshift within our range of validity. This is a practical, redshift‑dependent alternative to the common assumption of a constant spin distribution, and provides a useful input to semi-empirical and semi‑analytic models of galaxy formation.}

\end{abstract}

\section{Introduction}
\label{sec:introduction}

Within the hierarchical structure formation framework, dark matter halos serve as the building blocks in which the baryonic diffuse matter collapses to form galactic discs \citep[][]{White78,Fall80,Mo98,Zavala08}. There is an extensive literature on how dark matter halos gain their angular momentum through gravitational tidal interactions with the large scale mass distribution of the universe \citep[][]{Peebles69,Doroshhkevich70,Efstathiou80,White84}. This theory, known as Tidal Torque Theory, has been shown to be consistent with the results of cosmological N-body simulations and captures well the growth of the angular momentum of proto-halos \citep[][]{Barnes87,Sugerman00,Porciani02a,Porciani02b}. However, its validity breaks down in the non-linear regime, when halos have collapsed and virialised \citep[][]{White84,Barnes87,Porciani02b,Bailin05}, and this is understood to reflect the importance of merger and accretion in the assembly histories of halos \citep[][]{Gardner01,Maller02,Vitvitska02,benson.etal.2020}, and consequently the galaxies that they host. 

In this context, cosmological simulations provide crucial insight into how halos and galaxies acquire their angular momentum over time. The properties of galaxies seem to be strongly correlated to those of the dark matter halo in which they are embedded
\citep[][]{Mo98, Dutton07}. For instance, a halo's angular momentum can be connected to properties of galactic discs, such as their size and surface density \citep[][]{Fall80}, which in turn affects their star formation rates \citep[][]{vanderBosch00}. This relationship between halo angular momentum and the intrinsic properties of galaxies is one of the core assumptions underpinning semi-analytic models (SAMs) of galaxy formation, which evolve baryons in the background of dark matter halos by means of a coupled set of differential equations which parameterise key astrophysical processes, such as cooling, star formation, and stellar feedback \citep[][]{Kauffmann99a,Kauffmann99b,vanderBosch00,Croton06}. This approach has successfully reproduced many of the global properties of the galaxy population across cosmic time \citep[e.g.,][]{Knebe_eta_al_2015}. 



{Despite extensive work on the spin distribution at fixed epoch \citep[e.g.][]{Vitvitska02,HB06,Bett07,Maccio07,Maccio2008,Knebe08_spinmasscorr,MunozCuartas11,Ahn14,zjupa.and.springel.2017}, only a limited subset of studies has quantified its redshift evolution. In particular, \citet{HB06} present one of the few explicit evolutionary analyses - but, as we find in this paper, their results are affected by limited mass resolution\footnote{We reproduce their setup and show how resolution alters the inferred behaviour (\ref{app_sec:HB06}).}. Similarly, \citet{Ahn14} track $\lambda(z)$ and $\lambda^\prime(z)$ directly and find broadly consistent qualitative trends, but focus primarily on mass dependence and provide no analytical fitting relations for either the mean or the dispersion of the log-normal distribution as a function of redshift, making their results difficult to implement directly in models that require sampling from a log-normal at arbitrary $z$. \citet{zjupa.and.springel.2017} confirm the log-normal form of the spin distribution at multiple redshifts in the Illustris simulation (their Figures 9, 11, 18), but do not extract or model the redshift evolution of the distribution parameters quantitatively. Consequently, to our knowledge, no existing work provides closed-form, resolution-validated expressions for both $\mu(\ln\lambda, z)$
and $\sigma(z)$ simultaneously — the combination required to fully specify the spin distribution at any redshift for use in forward modelling. This is the gap our paper addresses.}

\medskip

The most common means to characterise a dark matter halo's angular {momentum} is the dimensionless spin parameter $\lambda$.
\citet{Peebles69} defined the spin parameter as,
\begin{equation} \label{eq:lambdaP}
\displaystyle
    \lambda = \frac{J \,|E|^{1/2}}{GM^{5/2}}
\end{equation}
\noindent where $G$ is Newton's gravitational constant; $J=\left|\sum_{i=0}^{N_p}\Vec{r_i}\times \Vec{p_i}\right|$ is the magnitude of the angular momentum of $N_p$ particles in the halo, each of mass $M/N_p$ where $M$ is the mass of the halo; and $|E| = |E_\text{kin} + E_\text{pot}|$ is the absolute value of the halo's total energy where $E_\text{kin}$ is its kinetic energy and $E_\text{pot}$ is its gravitational potential energy. We can interpret Equation~\ref{eq:lambdaP} as the halo's rotational velocity ($v_\text{rot}\sim\,J/MR$) normalised by its velocity dispersion ($\sigma\sim\,\sqrt{GM/R}\sim\sqrt{|E|/M}$). 

This classic \citet{Peebles69} definition has been supplemented subsequently by the \citet{Bullock01} definition,
\begin{equation} \label{eq:lambdaB}
\displaystyle
    \lambda' = \frac{1}{\sqrt{2}}\frac{J}{MVR} = \frac{1}{\sqrt{2}}\frac{J}{M \sqrt{GMR}}
\end{equation}
\noindent where $V=\sqrt{GM/R}$ is the circular velocity of the halo. We can interpret Equation~\ref{eq:lambdaB} as the halo's angular momentum ($J$) normalised by the angular momentum we would expect if halo material was on circular orbits ($\sim \sqrt{GMR}$). $\lambda'$ has the advantage that it is computationally cheaper to obtain than $\lambda$ because it avoids the expensive calculation of the gravitational potential energy, $E_\text{pot}$ \citep[][]{Vitvitska02,Knebe08_spinmasscorr}. Moreover, its value is less sensitive to the definition of a halo's boundary  \citep[][]{Bullock01}, which also makes it a versatile measure for the spin parameter of galaxies \citep[][]{Burkert04,Tonini06,Obreja22}.

There is a substantial body of work on the growth of the spin parameter of halos driven by the large-scale gravitational tidal field in the linear regime \citep[e.g.][]{Peebles69,Doroshhkevich70,White84,Catelan_and_Theuns_1996}. As halos enter the non-linear regime, their self-gravity dominates the large-scale tidal torques \citep[e.g.][]{Sugerman00} and merger and accretion events are predicted to become important \citep[e.g.][]{Maller02,Vitvitska02,Peirani_et_al_2004,HB06,Sharma_et_al_2012}{ - although the precise role of mergers in setting the spin distribution remains debated \citep[e.g.][]{Donghia_and_Navarro_2007,wang.and.white.2009}.}

\citet{Maller02} used a semi-analytic model to demonstrate that both tidal torques and orbital angular momentum transfer during mergers can produce similar halo spin distributions. \citet{Vitvitska02} used $N$-body simulations to model spin growth as a random walk driven by uncorrelated orbital angular momenta of merging subhalos. {This framework has been refined subsequently by \citet{benson.etal.2020}, who showed that accurately reproducing the N-body spin distribution requires including a weak correlation between the spins of progenitor halos and the angular momenta of their infalling subhalos, and by \citet{Hou.2025}, who used merger trees drawn from the Millennium and Millennium-II simulations to construct a stochastic model in which spin vector evolution is driven by the accretion of mass and angular momentum.}
\citet{Peirani_et_al_2004} found that halo angular momentum evolves in two distinct phases, with mergers increasing spin and smooth accretion reducing it, but despite these variations, the overall spin parameter distribution remains largely unchanged with redshift, apart from a slight increase in the mean value. \citet{Sharma_et_al_2012} showed that mergers redistribute angular momentum, causing the inner halo to lose spin while the outer regions gain it, leading to a net spin reduction of 40–80\% per Gyr. 

Interestingly, \citet{HB06} explicitly tracked the growth of both the Peebles ($\lambda$) and Bullock ($\lambda'$) spin parameters with redshift, and reported contrasting behaviour in their respective evolutions. They found that $\lambda'$ remains nearly constant but $\lambda$ increases with redshift. Evolution is driven by minor mergers and accretion rather than major mergers. Although major mergers boost $\lambda$ - by a factor of $\sim 1.3$ - their infrequency makes them a secondary influence. Similar trends were reported in \citet{Maccio07} and \citet{Ahn14}.

{It is worth noting that \citet{Donghia_and_Navarro_2007} argued that the apparent spin-up associated with major mergers is largely an artefact of unrelaxed systems, and that the virialisation process itself leads to a net decrease in spin through the ejection of high-angular-momentum material beyond the virial radius. More fundamentally, \citet{wang.and.white.2009} demonstrated that universal halo properties, including the spin distribution, are reproduced even in a hot dark matter universe where structure forms through monolithic collapse rather than hierarchical merging, suggesting that mergers may not be the necessary ingredient for establishing the log-normal form of the spin distribution.}


{Motivated by the contrasting findings in the literature, we revisit in this paper the evolution of the Peebles $\lambda$ and Bullock $\lambda^\prime$ spin parameters across redshift, with two specific aims: to provide a statistically robust, resolution-validated characterisation of how the mean and dispersion of the log-normal spin distribution evolve from $z=5$ to $z=0$; and to deliver closed-form fitting functions that allow physically motivated spin values to be drawn at any redshift within this range without recourse to simulations. Although previous studies have noted that $\lambda$ and $\lambda^\prime$ evolve differently \citep[e.g.][]{HB06,Ahn14}, no existing work provides analytical expressions for both the mean and the width of the distribution simultaneously — the combination required to fully specify the spin distribution at arbitrary redshift. We provide these here, and discuss the physical origin of the differences between the two definitions in Section~\ref{sec:conclusions}.}

The main focus of this work is to quantify, for a dark matter halo sample found in cosmological simulations, the redshift evolution of both spin definitions (Equations~\ref{eq:lambdaP} and \ref{eq:lambdaB}). The approach will be to study the statistical properties of the spin in the halo population. To this end, this work is structured as follows: in Section~\ref{sec:data} we introduce our suite of simulations, the halo finder used to identify dark matter halos, and provide a brief discussion on the spin and the halo sample's basic properties. Secondly, in Section~\ref{sec:spin_zevol} we showcase our main results, based upon studying in detail the redshift evolution of the spin probability distribution, eventually providing fitting functions that describe this evolution. Lastly, in Section~\ref{sec:conclusions} we present a discussion of our results and state the main takeaways of our experiments.

\section{Data \& Methodology}
\label{sec:data}

A detailed understanding and characterisation of the spin parameter of dark matter halos requires numerical simulations that accurately model the relevant physical processes - in this case, the gravitational collapse and dynamical evolution of halos in a cosmological background - with sufficient resolution to produce reliable results, and recover statistical samples of halos throughout a wide range of masses and scales; a halo finding code that correctly identifies and characterises the overdensities as bound objects; and appropriate post-processing pipelines for their analysis. In this section we introduce all these technical details and provide some basic discussion of the resulting dataset, setting the ground for the main results of our work.

\subsection{Simulations and Halo Finder}
\label{subsec:TheSims}

\noindent{{Simulation Details:}} We have run four dissipationless $\Lambda$-Cold Dark Matter ($\Lambda \unit{CDM}$) cosmological simulations using {\small{GADGET4}} \citep{GADGET4}. Initial conditions at redshift $z=63$ were created on-the-fly using the internal {\small{N-GENIC}} functionality in {\small{GADGET4}}, configured to generate and use a realisation of the \citet{Eisenstein1999} power spectrum of primordial density fluctuations. We stored 48 output snapshots between $z=5$ and $z=0$, uniformly distributed in $\log_{10}(a)$ space. The mass resolution of our runs is insufficient to explore higher redshifts reliably (i.e. $z\geq\,5$), but we consider the redshift range large enough for the conclusions drawn from our data.

\begin{table*}[t]
\centering
\resizebox{\textwidth}{!}{
\begin{tabular}{c|c c c|c c c c c c}
Simulation & $B \, [Mpc \, h^{-1}]$ & $m_{sim} \, [10^7 \cdot M_{\odot} \, h^{-1}]$ & $\epsilon \, [kpc \, h^{-1}]$ & $N_{h}(z=0.0)$ & $N_{h}(z=1.1)$ & $N_{h}(z=5.0)$ & $N_h^{f}(z=0.0)$ & $N_h^{f}(z=1.1)$ & $N_h^{f}(z=5.0)$\\
\hline
   B20 & $20$ & $1.7$ & $1.95$ & $19233$ & $22819$ & $11959$ & $1115$ & $1334$ & $267$\\
   B40 & $40$ & $3.4$ & $3.91$ & $22714$ & $25539$ & $8223$ & $1330$ & $1364$ & $81$\\
   B80 & $80$ & $6.8$ & $7.81$ & $26807$ & $27746$ & $3705$ & $1515$ & $1201$ & $8$\\
   B160 & $160$ & $13.6$ & $15.63$ & $31211$ & $27792$ & $614$ & $1511$ & $684$ & $0$\\
   Total & $-$ & $-$ & $-$ & $99965$ & $103896$ & $24501$ & $5471$ & $4583$ & $356$\\
\hline
\end{tabular}
}
\caption{Simulation parameters in the $B$-suite (columns 1-3), number of halos $N_h(z)$ found by AHF with more than 500 particles at redshifts $z = 0.0$, $1.1$ and $5.0$ (columns 4-6, respectively), and number of field halos $N_h^{f}(z)$ after additionally filtering out subhalos (columns 7-9).}
\label{tab:Bsuite}
\end{table*}

\smallskip

Each simulation follows the evolution of $N=256^3$ equal mass particles and vary only in the box size $B$, whose values are given in column 1 of Table~\ref{tab:Bsuite}. The particle masses varies as $m_{sim} = \rho_{crit} \Omega_0 \left(B/N\right)^3$, where $\rho_{crit} = 3H_0^2/8\pi G$, as shown in column 2 of Table~\ref{tab:Bsuite}. Even with this modest value of $N$ we can probe a wide range of halo masses from $z=5$ to the present day. We choose a gravitational softening of $\epsilon = 0.025\,B/N^{1/3}$, which is intermediate between the converged and optimal softenings of  \citet{Ludlow_et_al_2019}; the values used in each of the simulations is shown in Table~\ref{tab:Bsuite}. We adopt cosmological parameters of $\Omega_0 = 0.315$, $\Omega_{\Lambda}=0.685$, $h=67.4$ and $\sigma_8 = 0.810$ and $n_s=0.965$, in agreement with the latest observations \citep[][]{Planck2018}.


\medskip

{It is worth noting that, although there is an abundance of state-of-the-art simulation datasets publicly available, there are good reasons for running our own suite of simulations tailored to our specific needs. First, we require a uniform minimum particle number across the different box sizes to maintain a common resolution threshold over a broad halo‑mass range and redshift interval. Second, we need both spin definitions to be computed in a consistent manner - crucially, the Peebles spin needs gravitational potential energy, which is not generally available in public catalogues, and post‑processed potential energy calculations across dozens of snapshots and simulation suites are computationally prohibitive. Third, our science goal demands explicit resolution‑convergence tests and volume‑dependence checks, which we provide in the Appendix; this shows that resolution and box size influence the inferred $\lambda^\prime$ evolution, underscoring the need for controlled systematics.}

\medskip

\noindent{{Halo Finding:}} Dark matter halos were identified using the Amiga Halo Finder algorithm\footnote{Publicly available at \url{http://popia.ft.uam.es/AHF/}} \citep[AHF; see ][]{AHF}. The halo mass is determined by truncating the halo edge at a distance $R$ from the center, where the mean density within the halo volume goes below a predefined threshold of $\Delta(z)$ multiplied by a reference density $\rho_{ref}(z)$. For our study, we will focus on the $R_{200}$ definition, which uses $\Delta(z)=200$ and $\rho_{ref}(z) = \rho_{crit}(z)$, determining the halo masses as \citep[][]{Knebe11_MADHalos_ComparisonProj,Knebe13_HaloFinding_StateOfAffairs}
\begin{equation} \label{eq:halomass}
\displaystyle
    M_{200} = \frac{4}{3}\pi R_{200}^3 \cdot 200\rho_{crit}(z) .
\end{equation}
Note that we show in~\ref{app_sec:Rvir_results} our results for when $R_{vir}$ is defined by $\Delta(z)$ as provided by the spherical top-hat collapse model and $\rho_{ref}(z) = \Omega_m(z) \rho_{crit}(z)$ is the reference density \citep[cf.][]{Knebe13_HaloFinding_StateOfAffairs}. We find that while the results show only a minor quantitative dependence on the definition of the halo's boundary, the qualitative behaviour and key conclusions remain consistent with those presented in the main analysis.
    



\subsection{Halo Sample}
\label{subsec:HaloSample}

This section provides a succinct summary of the properties of our halo sample. We investigate three key aspects: first, we ensure that our sample spans a broad mass range with sufficient resolution (Section~\ref{subsubsec:MassFunc}); second, we assess the correlation between spin and mass and compare to previous studies (Section~\ref{subsubsec:spinmasscorrel}); and third, we characterise the spin probability distribution and compare with theoretical expectations (Section~\ref{subsubsec:SpinDist}). This last point is particularly important because the statistical properties of the spin distribution play a crucial role in our analysis. By addressing these aspects, we establish a solid framework for interpreting the results presented in Section~\ref{sec:spin_zevol}.

\subsubsection{Cumulative Halo Mass Function}
\label{subsubsec:MassFunc}

One of the main tools to study the consistency of a cosmological simulation and its mass range is the cumulative halo mass function $N(>M_{\rm 200})$, hereafter referred to as cHMF. 

\begin{figure}[httbp!]
\centering
\resizebox{\textwidth}{!}{
\includegraphics[width=7cm, trim = 15 0 10 0, clip]{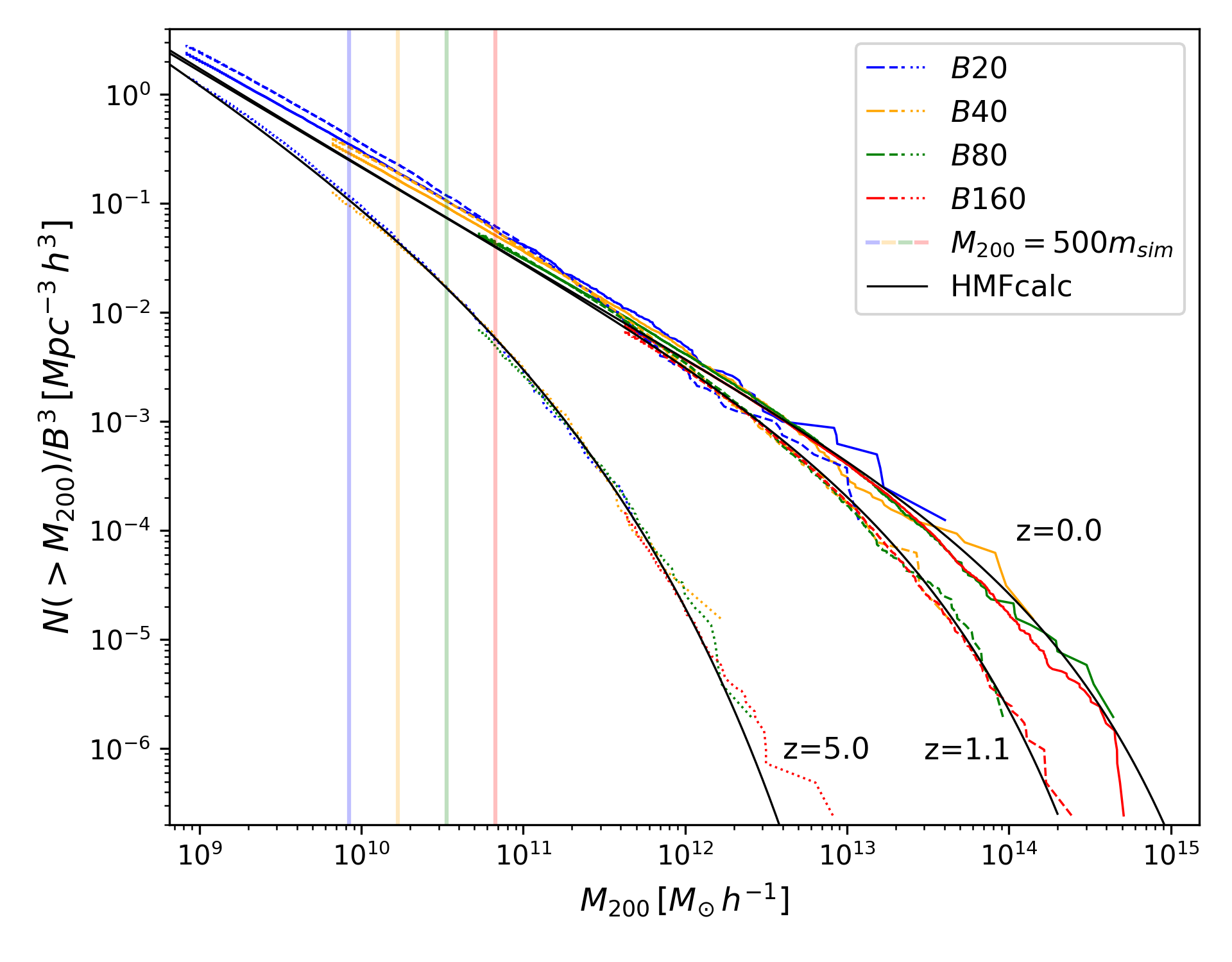}
\caption{Volume-normalised cumulative halo mass functions of our $N=256^3$ suite of simulations with $B=20$ (blue), $40$ (orange), $80$ (green) and $120 \, \unit{Mpc \, h^{-1}}$ (red) at redshifts $\unit{z = 0.0}$ (solid), $\unit{z = 1.1}$ (dashed) and $\unit{z = 5.0}$ (dotted). The solid black curves indicate the analytical cumulative mass distributions obtained using the \textsc{HMFcalc} code \citep[][]{HMFcalc} at each redshift, whereas the vertical lines indicate the mass of halos with 500 particles in each simulation, coloured accordingly. The leftmost end of the curves correspond to the minimum 20-particle limit for a halo.}
\label{fig:massfunc_R200}
}
\end{figure}

Figure~\ref{fig:massfunc_R200} shows the volume-normalised\footnote{The normalisation is done so as to compare simulations with varying volume $B^3$.} cHMF of the dark matter halos found in each of our simulations: blue, orange, green and red correspond, respectively, to B20, B40, B80 and B160 (as designated in Table~\ref{tab:Bsuite}), at redshifts $z=0.0$ (solid), $1.1$ (dashed curves), and $5.0$ (dotted). The theoretical results, calculated using the \textsc{HMFcalc} code \citep[][]{HMFcalc} at the corresponding redshifts, are shown as solid black curves. Furthermore, the mass of halos with 500 particles in each simulation is marked by vertical lines, coloured according to their corresponding simulation. This minimum particle number will be justified in more detail below, and it sets the mass limit of objects considered in our study.

The main point of Figure~\ref{fig:massfunc_R200} is to show that -- stacked together -- our suite of simulations covers a range of halo masses from approximately $10^{10}h^{-1}M_{\odot}$ to close to $10^{15}h^{-1}M_{\odot}$ - smaller simulation volumes (e.g. $B20$) provide insight into the low-mass halo regime, whereas larger volume (e.g. $B160$) probe the highest mass halos. The combination of the $B$-suite results in a continuous halo mass function which covers a wide mass range and follows well the \textsc{HMFcalc} theoretical expectations.
Note, however, that we apply two additional selection criteria.

\smallskip

\noindent{{Halos must contain at least 500 particles:}} This ensures sufficient resolution of a halo's internal properties to compute its spin parameter \citep[see, for instance, Appendix A in][for a study of the resolution effects on different halo measurements]{Allgood2006}\footnote{{We note that \citet{benson.2017} demonstrated that particle noise introduces a systematic upward bias in halo spin measurements in N-body simulations, with the noise-free median spin being approximately 3 per cent lower than the directly measured value, and that measuring an individual halo's spin to 10 per cent precision requires of order $4 \times 10^4$ particles. However, because our analysis is based on fits to the spin distribution or the halo population rather than measurements of individual halo spins, and because \citet{benson.2017} showed that the noise bias on the population median is small compared to the statistical uncertainties arising from halo counts, we do not expect this effect to significantly affect our conclusions.}}.

\smallskip

\noindent{{Substructure halos are removed:}} Tidal stripping of substructure and its gravitational interactions with the host halo affect its angular momentum and spin. In addition, unambiguous definition of material associated with the halo is difficult for substructure.  

\smallskip

\noindent For the remainder of this work, we combine halos from all simulations that satisfy these criteria into a single catalogue, which we summarise in Table~\ref{tab:Bsuite} (cf. $N_h^{f}$, columns 7-9).

\subsubsection{Spin-Mass Correlation}
\label{subsubsec:spinmasscorrel}
\begin{figure}[httbp!]
\centering
\resizebox{\textwidth}{!}{
\includegraphics[width=8cm, trim = 10 0 10 0, clip]{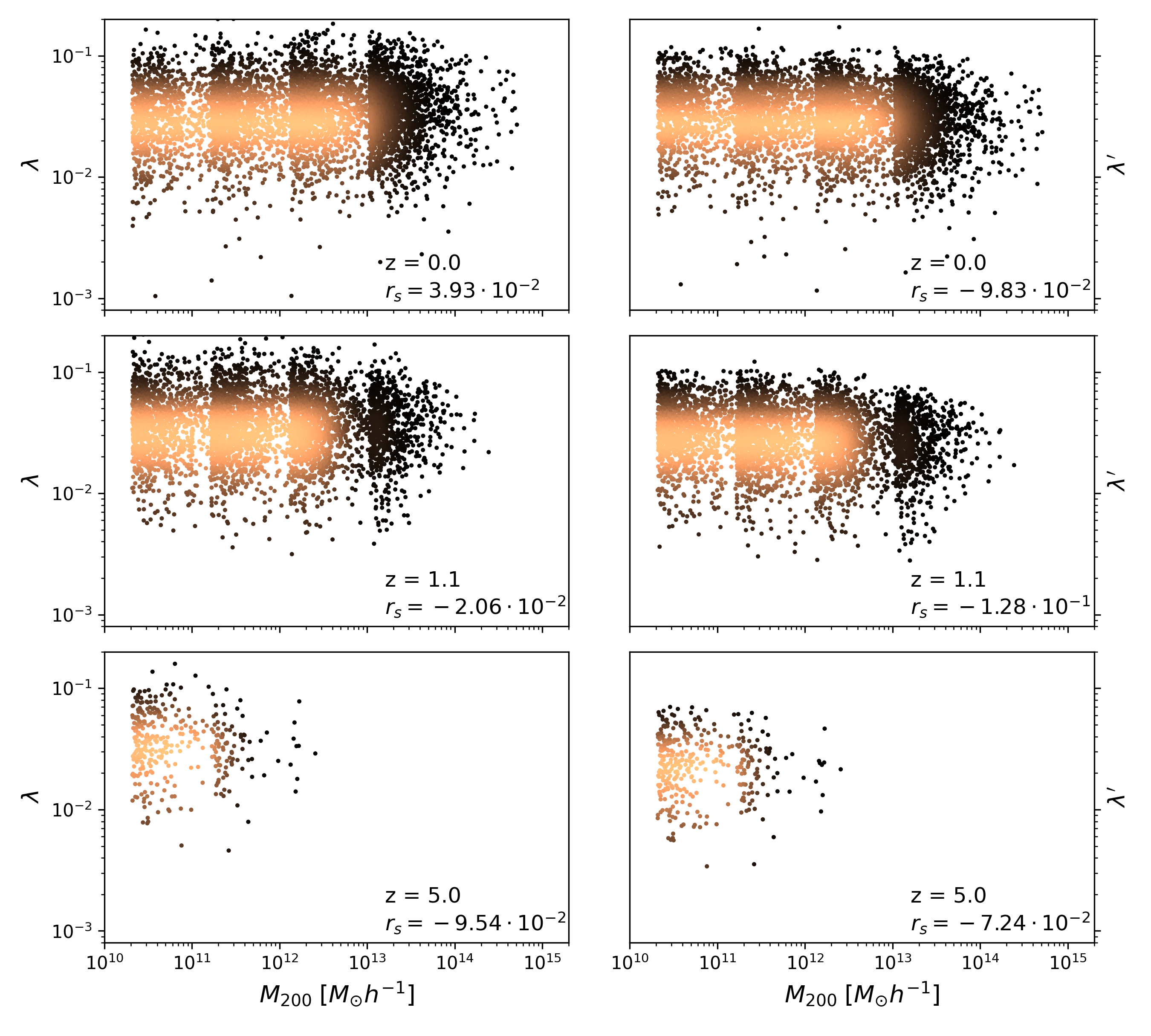}
\caption{Correlation between spin and mass at different redshifts $z$, coloured by density of data points. Each row indicates, from top to bottom respectively, redshifts $\unit{z} = 0.0$, $1.1$, and $5.0$, whereas left and right columns indicate the $\lambda$ and $\lambda '$ data, respectively. The text on the bottom right of each plot indicates the redshift $z$ and the spearman rank coefficient $\unit{r_s}$.}
\label{fig:spinvsM_R200}
}
\end{figure}
If we are to reliably identify redshift-dependent trends in the spin parameter, then we must account for a potential degeneracy with halo mass. This is because the maximum mass decreases with increasing redshift (see Figure~\ref{fig:massfunc_R200}), and so an observed evolution of spin with redshift may reflect the underlying trend with halo mass rather than a genuine evolutionary effect. To address this, in Figure~\ref{fig:spinvsM_R200} we investigate the correlation between the Peebles spin ($\lambda$), left panels) and Bullock spin ($\lambda'$, right panels) and halo mass at three representative redshifts - $z = 0$, $z = 1.1$, and $z = 5$ (top to bottom) - corresponding to those in Figure~\ref{fig:massfunc_R200}.  Each panel also includes the Spearman rank correlation coefficient for the halo sample at that redshift.

Figure~\ref{fig:spinvsM_R200} reveals that 
there is minimal correlation between spin and halo mass, as indicated by the low Spearman coefficients. The correlation is weakly negative, except for $\lambda$ at $z = 0$, and tends to strengthen with increasing redshift. This trend is consistent for $\lambda$, while for $\lambda'$, the strongest correlation appears at $z = 1.1$. As noted by \citet{Bett07}, the small magnitude and substantial scatter in $\lambda$ and $\lambda'$ makes this correlation difficult to detect. Our results are consistent with \citep[][]{Maccio07,Knebe08_spinmasscorr,MunozCuartas11,Ishiyama13}, which gives us confidence that any redshift evolution we observe in the mean value of the spin parameter is not driven by a correlation between spin and mass of the underlying halo population.

\subsubsection{Spin Distribution}
\label{subsubsec:SpinDist}
\begin{figure}[httbp!]
\centering
\resizebox{\textwidth}{!}{
\includegraphics[width=6.5cm]{plots/lambda_R200_N256_rawdist_log_host_revised.png}
\caption{Probability distribution of $\ln (\lambda)$ (top panel) and $\ln \lambda'$ (bottom panel) for our $N=256^3$ $B$-suite of simulations at redshifts $\unit{z} = 0.0$ (black filled squares), $\unit{z} = 1.1$ (red filled squares) and $\unit{z} = 5.0$ (magenta filled squares), all using Poisson uncertainty. The solid, dashed and dotted curves are Gaussian fits to the distributions at each redshift discussed respectively, coloured according to their associated dataset. The coloured vertical lines show the mean values of their same-coloured dataset at each redshift.}
\label{fig:spindist_log_R200}
}
\end{figure}

There is a rich body of literature \citep[e.g.,][]{Efstathiou87,Bullock01,Vitvitska02,Maller02,HB06,Bett07,Knebe08_spinmasscorr,AntonuccioDelogu10,Davis10,MunozCuartas11,Ahn14,zjupa.and.springel.2017} that has demonstrated that the spin probability distribution is well fit by a log-normal function{, which can be understood physically as a consequence of the stochastic nature of angular momentum growth\footnote{For example, \citet{kim.etal.2015} have shown that log-normality of the distribution arises directly from a geometric Brownian motion model for halo spin evolution driven by mergers and accretion.}.}
We verify this
in Figure~\ref{fig:spindist_log_R200}, in which we show the probability distribution of $\ln(\lambda)$ (top panel) and $\ln(\lambda')$ (bottom panel) at redshifts $z=0.0$, $1.1$ and $5.0$ (black, red and magenta, respectively). 

Here the uncertainties are calculated using Poisson statistics. 
Both panels also show Gaussian fits to each dataset as continuous, dashed and dotted curves respectively (colour-coded by redshift). Vertical lines indicate the mean value of the data at each redshift, for comparison with the Gaussian fits. In all subsequent analysis we use the mean and standard deviation as derived directly from the data rather than from the Gaussian fits.

Figure~\ref{fig:spindist_log_R200} confirms that both $\ln(\lambda)$ and $\ln(\lambda')$ are well fit by Gaussians, especially at low redshifts. There are small offsets between mean values measured from the data (marked by the vertical lines) and the values at which the Gaussians peak, which we would expect to be smaller for larger halo samples. At higher redshifts, the larger offsets could be an effect of the binning or an intrinsic property of the data itself, but there is still very good consistency between the data and the fits even in this regime.

\section{The Redshift Evolution of the Spin Parameter}
\label{sec:spin_zevol}
In this section, we present our results for the redshift evolution of both the mean (Sec.~\ref{subsec:mean_spin_z}) and standard deviation  (Sec.~\ref{subsec:std_spin_z}) of $\ln(\lambda)$ and $\ln(\lambda')$. We define the probability distribution of the mean by, 
\begin{equation}
    \mathcal{L}^{(\prime)}(z) = \frac{1}{N_h^{f}(z)} \sum_{i=1}^{N_h^{f}(z)} \ln \left[\lambda_i ^{(\prime)} (z) \right],
\end{equation}
and of the standard deviation by,
\begin{equation}
    \mathcal{S}^{(\prime)}(z) = \sqrt{\frac{1}{N_h^{f}(z)-1} \sum_{i=1}^{N_h^{f}(z)} \left| \mathcal{L}^{(\prime)}(z)-\ln \left[ \lambda_i^{(\prime)}(z) \right] \right|^2}.
\end{equation}
Here the 'primed' notation indicates whether we are dealing with $\lambda$ or $\lambda'$, and $N_h^{f}(z)$ is the number of field halos at a given redshift. We also demonstrate that our convergence criterion of 500 is sufficient.
 
\subsection{Mean value $\mathcal{L}$}
\label{subsec:mean_spin_z}

\begin{figure}
\centering
\resizebox{\textwidth}{!}{
\includegraphics[width=6cm, trim = 10 0 5 0, clip]{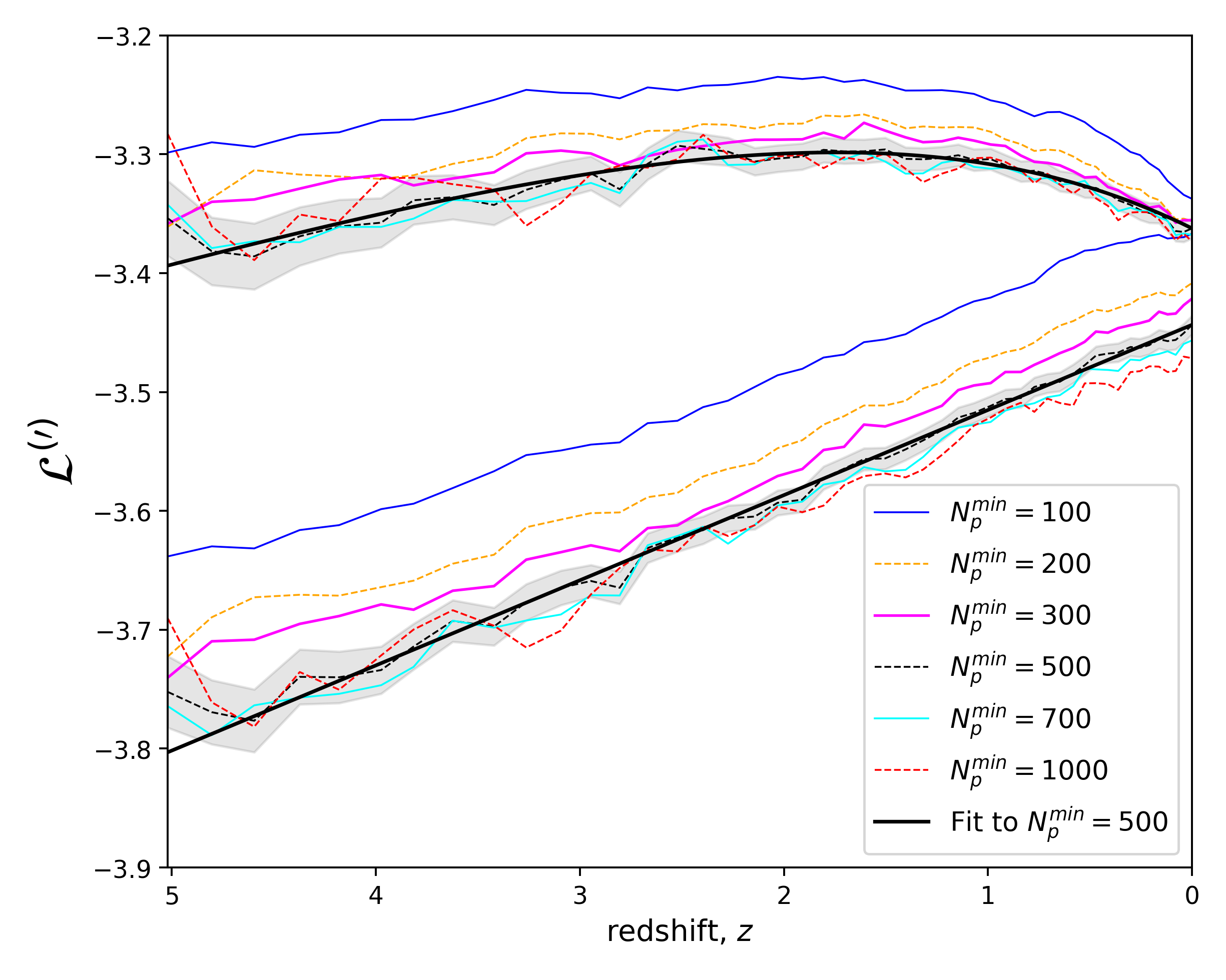}
\caption{Redshift evolution of $\mathcal{L}(z)$ and $\mathcal{L}'(z)$, for our $N=256^3$ $B$-suite of simulations using different $N_{p}^{min}$ filtering in the range $[100,1000]$ (blue, magenta and cyan solid curves, and orange, black and red dashed curves; see legend). The Poisson uncertainty of the $N_p^{min}=500$ data is shown as shaded gray regions, and Equations~\ref{eq:meanspinP_fit} and \ref{eq:meanspinB_fit} fit to $\mathcal{L}(z)$ and $\mathcal{L}'(z)$ respectively are plotted as solid black curves. The Figure includes two separate datasets displaying the same style but different behaviour: the bottom set corresponds to $\mathcal{L}(z)$, whereas the top set corresponds to $\mathcal{L'}(z)$. Using only host halos.}
\label{fig:meanlog_Ncuts_R200}
}
\end{figure}

\paragraph*{{Redshift evolution:}} Figure~\ref{fig:meanlog_Ncuts_R200} shows $\mathcal{L}^{\prime}(z)$ (upper set of curves) and $\mathcal{L}(z)$ (lower set of curves) plotted against redshift, $z$. The different colours correspond to different minimum number of particles per halo, with $N_h^{f}(z) \in [100, 1000]$ (see caption for further details). 

The uncertainties on $\mathcal{L}(z)$, which we compute as 
\[\mathcal{S}^{(\prime)}(z)\displaystyle\ /\ \sqrt{N_h^{f}(z)},\]
are shown as shaded grey regions, only for the datasets with $N_p^{\text{min}} = 500$. For these same datasets, we also include fits to analytical functions - described below - shown as a continuous thick black line.

We observe a striking contrast between the evolution of the mean values estimated from the logarithm of the spin parameter definitions of $\ln(\lambda)$ and  $\ln(\lambda')$. The mean $\ln(\lambda)$ increases approximately linearly with decreasing redshift - consistent with previous findings \citep[e.g.,][]{HB06} -  while the mean $\ln(\lambda')$ increases gradually with decreasing redshift to $z \sim 2$, after which it begins to decline - which is inconsistent with \citet{HB06}. However, as we demonstrate in~\ref{app_sec:HB06}, this inconsistency in $\lambda'$ is an artifact of mass resolution and simulation volume. If we perform simulations that match those of \citet{HB06}, then we recover their results. This gives us confidence that we understand this discrepancy, and that our choice of a minimum number threshold of 500 particles is sufficient for convergence.

We note that \citet[][]{Ahn14} have studied the redshift evolution of the \citet{Peebles69} and \citet{Bullock01} spins, as well as a modified form (an empirical, redshift-dependent, correction) and find that spin decreases with increasing redshift (cf. their Figure 12).
These authors track the evolution of $\lambda$ and $\lambda'$ rather than $\ln(\lambda)$ and $\ln(\lambda')$, and so a like-for-like comparison is not straightforward, but it is noteworthy that they report a steady monotonic increase with decreasing redshift in $\lambda$, whereas $\lambda'$ increases monotonically with decreasing redshift before it plateaus at $z\simeq1$ at lower halo masses. This is broadly consistent with our results in Figure~\ref{fig:meanlog_Ncuts_R200}.

\medskip

\paragraph*{{Fitting functions:}} The fitting functions used to model the redshift evolution of the mean logarithmic spin parameter are shown in Figure~\ref{fig:meanlog_Ncuts_R200} as solid black curves. These fits were applied to the $N_p^{\rm min} = 500$ dataset (represented by the dashed black curves), ensuring the stability and reliability of the derived parameters.

For the mean \citet{Peebles69} spin parameter, $\mathcal{L}(z)$, we adopt a simple linear model:
\begin{equation}
\label{eq:meanspinP_fit}
    \mathcal{L}(z) = \mathcal{L}(0) + l \, z \ ,
\end{equation}
\noindent where $l$ is the sole free parameter, and $\mathcal{L}(0)$ is fixed by the simulation data. This choice reflects our primary interest in the evolutionary trend of the spin parameter rather than its absolute normalisation.

In contrast, the mean \citet{Bullock01} spin parameter, $\mathcal{L}'(z)$  exhibits a non-monotonic behaviour, with a turnover at lower redshifts, $z\simeq1$. To capture this feature, we extend the linear model by introducing a logarithmic correction term:
\begin{equation}
\label{eq:meanspinB_fit}
    \mathcal{L}'(z) = \mathcal{L}'(0) + l' \, z + \ln \left( \frac{1 + e^{ab}}{1 + e^{-a(z - b)}} \right) \ ,
\end{equation}
\noindent where $l'$, $a$, and $b$ are free parameters. The logarithmic correction term in Equation~\ref{eq:meanspinB_fit} captures the turnover behaviour of $\lambda'$ at low redshift, allowing the model to recover the simulation-derived value at  $z$ = 0 and accurately trace its non-linear evolution.

Both fitting functions are well matched to the simulation data, as shown in Figure~\ref{fig:meanlog_Ncuts_R200}. The corresponding fit parameters are listed in Table~\ref{tab:fitparams}. Note that the validity of these fits has been tested only within the redshift range $0.0 \leq z \leq 5.0$ and for halo masses in the range $9.9 \leq \log_{10}(M / M_{\odot} h^{-1}) \leq 15.3$. There are relatively large uncertainties in some of the parameters, but this could be reduced with a larger halo sample. 

{The absolute change in $\mathcal{L}(z)$ is modest compared to the intrinsic scatter ($\approx 0.5$), but it is measured with high statistical precision. The cumulative shift between $z=5$ and $z=0$ is $\approx 0.3$ in $\ln\lambda$, or about 0.7 of a standard deviation. We describe the evolution as measurable and statistically robust rather than large in amplitude.}

\noindent{{Sensitivity of results to halo definition:}} We have assessed the extent to which these results are sensitive to mass resolution and our choice of halo definition. We have already noted above the effects of mass resolution in our comparison to \citet{HB06} (see~\ref{app_sec:HB06}). Our choice of halo definition could lead to potential pseudo-evolution, which arises because halos are defined relative to a background density which is a function of redshift  \citep[cf.][]{Diemer2013}. We investigate this by using $M_{\rm vir}$ rather than $M_{200}$ as halo mass definition (see~\ref{app_sec:Rvir_results}) and show that the qualitative trends are consistent even though the precise values of the fit parameters differ. These tests confirm that our fitting functions reliably describe the spin evolution, although the fit parameters themselves depend on halo definition.

\subsection{Standard deviation $\mathcal{S}$}
\label{subsec:std_spin_z}

\begin{figure}
\centering
\resizebox{\textwidth}{!}{
\includegraphics[width=7cm, trim = 10 0 5 0, clip]{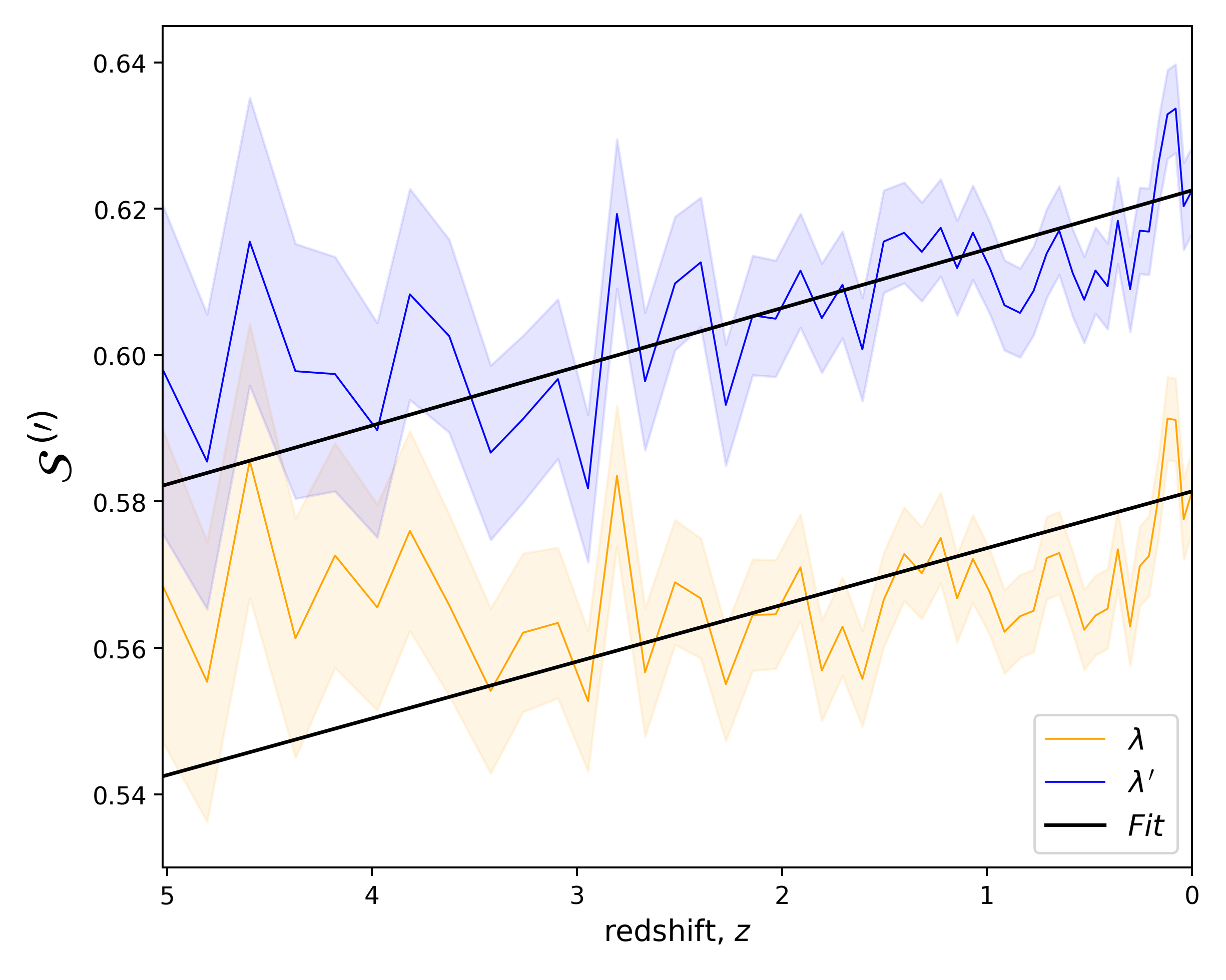}
\caption{Redshift evolution of $\mathcal{S}(z)$ (orange curve) and $\mathcal{S'}(z)$ (blue curve) respectively, for our $N=256^3$ $B$-suite of simulations, with Poisson uncertainty marked by shaded regions correspondingly coloured, and Equation~\ref{eq:meanspinP_fit} fit to the data as black solid curves. Using only host halos with $N_{p} \geq 500$.}
\label{fig:stdlog_Ncuts_R200}
}
\end{figure}

\paragraph*{{Redshift evolution:}}
Figure~\ref{fig:stdlog_Ncuts_R200} shows how the standard deviations in $\ln(\lambda')$, $\mathcal{S}^{\prime}(z)$ (upper curves), and $\ln(\lambda)$, $\mathcal{S}(z)$ (lower curves), vary with redshift, $z$ for field halos with more than 500 particles. The shaded regions correspond to the uncertainties, which we calculate as
\[\frac{\mathcal{S}^{(\prime)}(z)}{\sqrt{2 \left(N_h^{f}(z)-1\right)}} .\]
These uncertainties represent sampling errors on the measured mean and dispersion of $\ln\lambda$ and $\ln\lambda^\prime$ arising from finite halo counts.

The behaviour of $\mathcal{S}(z)$ and $\mathcal{S}'(z)$ are similar, both showing a weak linear increase with decreasing redshift. The normalisation of $\mathcal{S}'(z)$ is higher and it has a steeper slope, compared to $\mathcal{S}(z)$. This is consistent with \citet[][]{Ahn14}, who reported a positive correlation between the average spin and its dispersion. 


\medskip

\paragraph*{{Fitting functions:}}
The fitting functions used to model the redshift evolution of the standard deviations in the logarithmic spin parameters are shown in Figure~\ref{fig:stdlog_Ncuts_R200} as solid black curves. We use the same functional form as in the case of $\mathcal{L}(z)$ (Equation~\ref{eq:meanspinP_fit}) to describe $\mathcal{S}(z)$ and $\mathcal{S'}(z)$,
\begin{equation}
\label{eq:std_fit}
    \displaystyle
\mathcal{S}^{(\prime)}(z) = \mathcal{S}^{(\prime)}(0) + s^{\,(\prime)} \ z
\end{equation}
\noindent where $s^{\,(\prime)}$ is a free parameter of the model. We note that we have experimented with fitting functions that incorporate a degree of curvature, but we find that the linear fit is sufficiently accurate to not warrant the greater complexity and number of free parameters. Fit parameters are given in Table~\ref{tab:fitparams}. As for the fits to $\mathcal{L}(z)$ and $\mathcal{L'}(z)$, so too for $\mathcal{S}(z)$ and $\mathcal{S'}(z)$ - the validity of these fits has been tested only within the redshift range $0.0 \leq z \leq 5.0$ and for halo masses in the range $9.9 \leq \log_{10}(M / M_{\odot} h^{-1}) \leq 15.3$. Parameter uncertainties could be reduced with a larger halo sample.


\begin{table}[h]
\centering
\resizebox{\textwidth}{!}{
    \begin{tabular}{c|c c c | c}
     & \multicolumn{3}{c|}{$\mathcal{L}^{(\prime)}(z)$} & $\mathcal{S}^{(\prime)}(z)$\\
     \hline
     \hline
     & $10^2 \ l^{\ (\prime)}$ & $10 \ a$ & $b$ & $10^3 \ s^{\ (\prime)}$ \\
    \hline
    $\lambda$ & $-7.164 \pm 0.095$ & $-$ & $-$ & $-7.755 \pm 0.672$\\
    $\lambda'$ & $-5.093 \pm 1.520$ & $6.494 \pm 1.444$ & $-2.050 \pm 0.138$ & $-8.037 \pm 0.716$\\
    \hline
    \end{tabular}
    }
\caption{Fit parameters of $\mathcal{L}^{(\prime)}(z)$ (columns 1-3) to Equations~\ref{eq:meanspinP_fit} and \ref{eq:meanspinB_fit}, and of $\mathcal{S}^{(\prime)}(z)$ (column 4) to Equation~\ref{eq:std_fit}, for our $N=256^3$ $B$-suite of simulations. The fits were performed on the $N_p^{min}=500$ host halo dataset.}
\label{tab:fitparams}
\end{table}

\section{Discussion and conclusions}
\label{sec:conclusions}

This study has investigated the redshift evolution of the dark matter halo spin parameter in a $\Lambda$CDM cosmology, using a suite of $N$-body simulations with varying box sizes and a consistent halo selection criterion ($\geq 500$ particles, excluding substructure). We studied the two most commonly used spin parameters in the literature - those of \citet{Peebles69}, $\lambda$, and \citet{Bullock01}, $\lambda'$ - and provided a detailed statistical characterisation of their behaviour over the redshift range $0 \leq z \lesssim 5$. 

As a first step, we verified that there is minimal correlation between spin and halo mass, and that the spin distributions at all redshifts are well described by log-normal functions, in agreement with previous studies. This was important to ensure that the observed redshift evolution is not a byproduct of mass-dependent selection effects, and to validate our use of Gaussian statistics in our fitting of the data.

Our main results can be summarised as follows.
\begin{enumerate}

\item The mean of  $\ln(\lambda)$ increases approximately linearly with redshift, while $\ln(\lambda')$ shows non-monotonic behaviour — increasing from $z$=5 to $z\sim1-2$ before declining to $z$=0. These trends are robust across simulation volumes and halo definitions. We have compared our results with those of earlier studies \citep[e.g.][]{HB06,Ahn14} and demonstrated that, where there are inconsistencies (e.g. in evolution of $\lambda'$), accounting for mass resolution can reconcile differences.

\item The dispersion of spin values evolve with redshift. Both $\ln(\lambda)$ and $\ln(\lambda')$ show increasing standard deviation over time, as we would expect given the growing complexity of halo assembly histories due to mergers and accretion. However, $\ln(\lambda')$ consistently displays higher dispersion, suggesting it is more sensitive to dynamical processes, albeit more stable against the halo edge definition as discussed in Section~\ref{sec:introduction}.

\end{enumerate}

{The contrasting evolutionary behaviours of $\lambda$ and $\lambda^\prime$ can be understood physically by examining how their definitions respond to the evolving internal structure of halos. The Peebles $\lambda$ involves the total energy $|E|$, which is sensitive to the gravitational potential and hence to the halo concentration $c$. The Bullock $\lambda^\prime$ instead normalises by the circular velocity $V=\sqrt{GM/R}$ at the halo boundary, making it less sensitive to the internal mass distribution. Following \citet{Mo98}, the two definitions are approximately related by $\lambda / \lambda' \approx \sqrt{2}(M_\mathrm{vir}/M_{200})^{1/3} f(c)$, where $f(c)$ encodes the concentration dependence. Since halo concentrations are known to increase with decreasing redshift \citep{Maccio07, MunozCuartas11}, this predicts that $\lambda$ and $\lambda^\prime$ should evolve differently — with $\lambda$ more sensitive to the secular growth of concentration over time. This is consistent with our finding that $\mathcal{L}(z)$ evolves linearly while $\mathcal{L}'(z)$ shows a non-monotonic turnover near $z \simeq 1$, where concentration growth slows. The results of~\ref{app_sec:Rvir_results} support this interpretation - switching from $R_{200}$ to $R_\mathrm{vir}$ alters the effective overdensity threshold and hence the inferred concentration, and changes the quantitative fit parameters while leaving the qualitative trends intact; this indicates that the differential evolution is physical rather than a numerical artefact.}

We provide three fitting functions (Equations~\ref{eq:meanspinP_fit},~\ref{eq:meanspinB_fit}, and~\ref{eq:std_fit}) that accurately describe the redshift evolution of the mean and standard deviation of $\lambda$ and $\lambda^\prime$. {Our fitting functions offer a practical, physically motivated, and computationally inexpensive way to include a redshift‑dependent spin prescription into semi-empirical \citep[e.g.][]{behroozi.etal.2022} and SAMs \citep[e.g.][]{Lagos18}, as well as general theoretical modelling. We note that our N-body simulations, and consequently our fitting functions, do not account for the presence of baryons and galaxy formation processes (e.g. radiative cooling) in halo assembly, and so the trends we have measured will be modified when baryons are accounted for, as in hydrodynamical galaxy formation simulations. \citet{zjupa.and.springel.2017} have demonstrated that there are strong trends with mass and redshift in the spins of gas and stars within halos, which also impacts the inner dynamical structure of the halos. For this reason, we caution that the fitting functions we provide apply only to the dark matter component of the halo spin, and that additional modelling of baryonic effects will be required before they can be fully applied to the baryonic components in SAMs.}

{However, this does not diminish the practical value of our fitting functions for several reasons. First, even if individual halo and galaxy spins are only weakly correlated, halo spin enters SAMs through the specific angular momentum budget available for disc formation, setting the predicted disc scale lengths and structural properties. Second, the weak correlation measured in full-physics simulations applies primarily at low redshift, where our results show the evolutionary offset is modest; at $z > 2$, where disc formation and angular momentum transfer mechanisms are less well understood, the halo-galaxy spin connection remains poorly constrained. Third, regardless of the strength of the halo-galaxy spin connection, providing accurate, resolution-validated evolutionary tracks for a fundamental halo property — where none previously existed — is a worthwhile contribution to the community. Our fitting functions impose no additional computational cost and can trivially replace the common assumption of a constant spin distribution.}

\smallskip

From a theoretical perspective, the linear growth and lower dispersion of $\ln(\lambda)$ suggest it may serve as a more stable tracer of halo angular momentum evolution. However, the more complex evolution of $\ln(\lambda')$ and its larger dispersion means that trends in spin should show more variation and greater sensitivity to halo dynamics, and it should be more accessible to observations, given its definition (dependence on $V$ in Equation~\ref{eq:lambdaB} rather than $E$ in Equation~\ref{eq:lambdaP}); see, for instance, \citet[][]{Obreja22}

The differences between the Peebles and Bullock spin parameters, $\lambda$ and $\lambda'$, though explored in only a limited number of previous studies \citep[][]{HB06,Ahn14}, are shown in this work to be both significant and persistent across redshift. While both are proxies for halo rotational support, they encapsulate fundamentally different physical assumptions—particularly in their treatment of energy and angular momentum. These differences manifest in distinct evolutionary trends and statistical properties, reinforcing the need for careful consideration when applying spin parameters in theoretical and observational contexts. Future work will aim to disentangle the contributions of these assumptions and assess their impact on galaxy formation models.

\begin{acknowledgement}
{We thank the anonymous referee for thoughtful feedback that has helped to refine and sharpen the presentation of our results.} Part of this work was performed on the OzSTAR national facility at Swinburne University of Technology. The OzSTAR program receives funding in part from the Astronomy National Collaborative Research Infrastructure Strategy (NCRIS) allocation provided by the Australian Government, and from the Victorian Higher Education State Investment Fund (VHESIF) provided by the Victorian Government.
\end{acknowledgement}

\paragraph{Funding Statement}
AK is supported by the Ministerio de Ciencia e Innovaci\'{o}n (MICINN) under research grants PID2021-122603NB-C21 and PID2024-156100NB-C21. He further thanks Fugazi for `waiting room'. CP acknowledges the support of the Australian Research Council (ARC) Centre of Excellence for All Sky Astrophysics in 3 Dimensions (ASTRO 3D), through project number CE170100013. RMP acknowledges the support of the Australian Government through the Australian Research Council Centre of Excellence for Dark Matter Particle Physics (CDMPP, CE200100008). AU acknowledges the support of an Australian Government Research Training Program Scholarship. 

\paragraph{Competing Interests}
None

\paragraph{Data Availability Statement}
All data used in our analysis are available upon request.

\printendnotes

\printbibliography
\appendix

\section{Virial Radius Results}
\label{app_sec:Rvir_results}

To study the effect of halo definition on our results, we have repeated the analysis of Section~\ref{sec:spin_zevol} using the virial radius, $R_{vir}$, instead of $R_{200}$ as our halo boundary. Here we assume a virial overdensity that is redshift dependent \citep[cf.][]{Bryan98}, 
\begin{equation}
    \Delta_{vir}(z) = 18\pi^2 + 82f(z) - 39f(z)^2
\end{equation}
where
\begin{equation}
    f(z)=\frac{\Omega_m(z)(1+z)^3}{\Omega_m(z)(1+z)^3 + \Omega_{\Lambda}(z)} - 1
\end{equation}
\noindent where the reference density is the background density of the universe, $\rho_{ref}(z) = \Omega(z) \rho_{crit}(z)$. The masses are therefore determined by:
\begin{equation}
    M_{vir}=\frac{4}{3}\pi R_{vir}^3 \cdot \Delta(z)\Omega(z) \rho_{crit}(z)
\end{equation}
We expect spin values to be affected by this change in halo definition, but for the effect to be small. This should provide a test of stability of our results because they should not depend strongly on how the halo boundary is defined.

\begin{figure}
\centering
\includegraphics[width=7cm, trim = 10 0 5 0, clip]{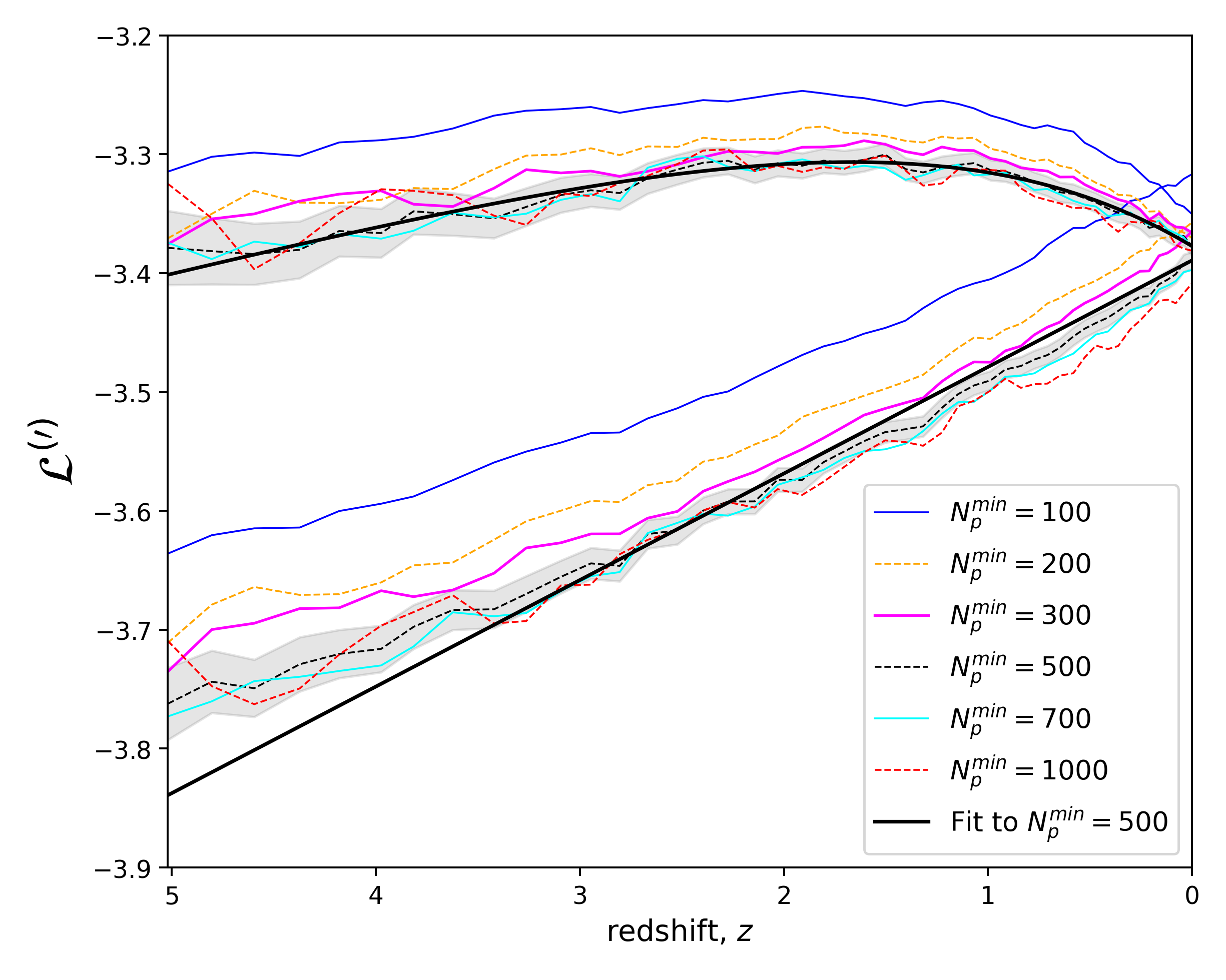}\\
\includegraphics[width=7cm, trim = 10 0 5 0, clip]{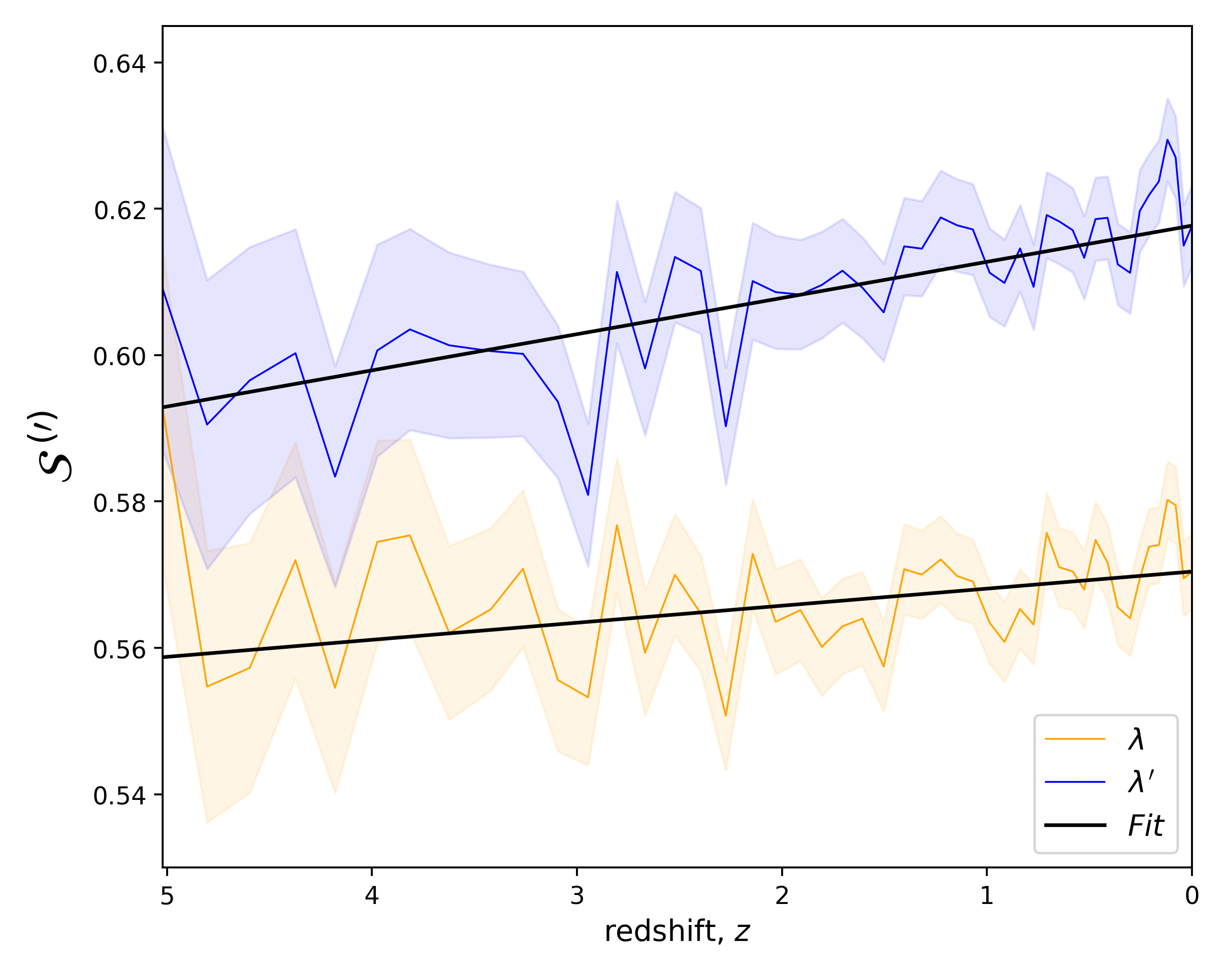}
\caption{Analog to Figures~\ref{fig:meanlog_Ncuts_R200} (top panel) and \ref{fig:stdlog_Ncuts_R200} (bottom panel) using the $R_{vir}$ halo edge definition.}
\label{fig:results_Rvir}
\end{figure}

Our main results were the redshift and $N_p^{min}$ dependence of $\mathcal{L}^{(\prime)}(z)$, and the redshift dependence of $\mathcal{S}^{(\prime)}(z)$. In Figure~\ref{fig:results_Rvir} we show the evolution of $\mathcal{L}(z)$ and $\mathcal{L'}(z)$ (upper panel) and $\mathcal{S}(z)$ and $\mathcal{S'}(z)$ (lower panel), following the conventions of Figures~\ref{fig:meanlog_Ncuts_R200} and \ref{fig:stdlog_Ncuts_R200}. We provide all fitting parameters resulting from our analysis in Table~\ref{tab:fitparams_Rvir}.

There is very good qualitative agreement between these results for $R_\text{vir}$ and those for $R_{200}$. The $\lambda$ values tend to be slightly larger and the curves are weakly convex, whereas the $\lambda '$ values tend to be slightly lower with no change in shape. The fitting functions provided by Equations~\ref{eq:meanspinP_fit} and \ref{eq:meanspinB_fit} work well for both spin definitions. There now exists a small deviation between the $\lambda$ values and the linear fit, although it is within the data's uncertainty.


In the bottom panel we see the effect of using $R_{vir}$ is mild on $\mathcal{S}^{(\prime)}(z)$. The quantitative values are very similar, with the largest difference occurring for $\lambda$, for which the evolution is now very weak. 

If we compare the fitting parameters of Tables~\ref{tab:fitparams} and \ref{tab:fitparams_Rvir}, we observe the differences are small. The largest differences occur for $\mathcal{S}^{(\prime)}(z)$, however the general magnitude of the values is within a reasonable range. Our results are therefore robust to the choice for the halo boundary.

\begin{table}[h]
\centering
\resizebox{\textwidth}{!}{
    \begin{tabular}{c|c c c | c}
     & \multicolumn{3}{c|}{$\mathcal{L}^{(\prime)}(z)$} & $\mathcal{S}^{(\prime)}(z)$\\
     \hline
     \hline
     & $10^2 \ m^{(\prime)}$ & $10 \ a$ & $b$ & $10^3 \ l^{(\prime)}$ \\
    \hline
    $\lambda$ & $-8.964 \pm 0.092$ & $-$ & $-$ & $-2.330 \pm 0.649$\\
    $\lambda'$ & $-4.362 \pm 1.025$ & $8.128 \pm 1.463$ & $-1.865 \pm 0.079$ & $-4.944 \pm 0.693$\\
    \hline
    \end{tabular}
    }
\caption{Analog to Table~\ref{tab:Bsuite} in the case of $R_{vir}$.}
\label{tab:fitparams_Rvir}
\end{table}

\section{Reproduction of Hetznecker \& Burkert Results}
\label{app_sec:HB06}

One of the earliest comparative studies of the spin parameters defined by \citet{Peebles69} and \citet{Bullock01} was conducted by \citet{HB06}. Given the relevance of their findings, it was important for us to verify whether our simulations could reproduce their results before proceeding with our own analysis. \citet{HB06} employed an ensemble of 82 cosmological $N$-body simulations with differing realisations of the initial power spectrum; $N=64^3$ particles; $B=20,\unit{Mpc,h^{-1}}$; and cosmological values $\Omega_0=0.3$, $\Omega_{\Lambda}=0.7$, $h=65$, and $\sigma_8=1.13$. Their analysis was restricted to halos containing more than 500 particles.

To replicate their setup, we ran two additional low-resolution simulation ensembles using $B=20,\unit{Mpc,h^{-1}}$ and the cosmology described in Section~\ref{sec:data}. The first ensemble consisted of 40 runs with differing realisations of the initial power spectrum with $N=64^3$ particles, and the second comprised 10 different realisations with $N=128^3$. Our goal was to reproduce the results of \citet{HB06} using the lowest-resolution set, and to observe convergence toward the results presented in the main body of this work as resolution increased. For this purpose, we treat our $B$-suite simulations as two distinct cases: the $B=20,\unit{Mpc,h^{-1}}$ simulation in isolation, and the full $B$-suite.

\begin{figure}
\centering
\resizebox{\textwidth}{!}{
\includegraphics[width=6.5cm]{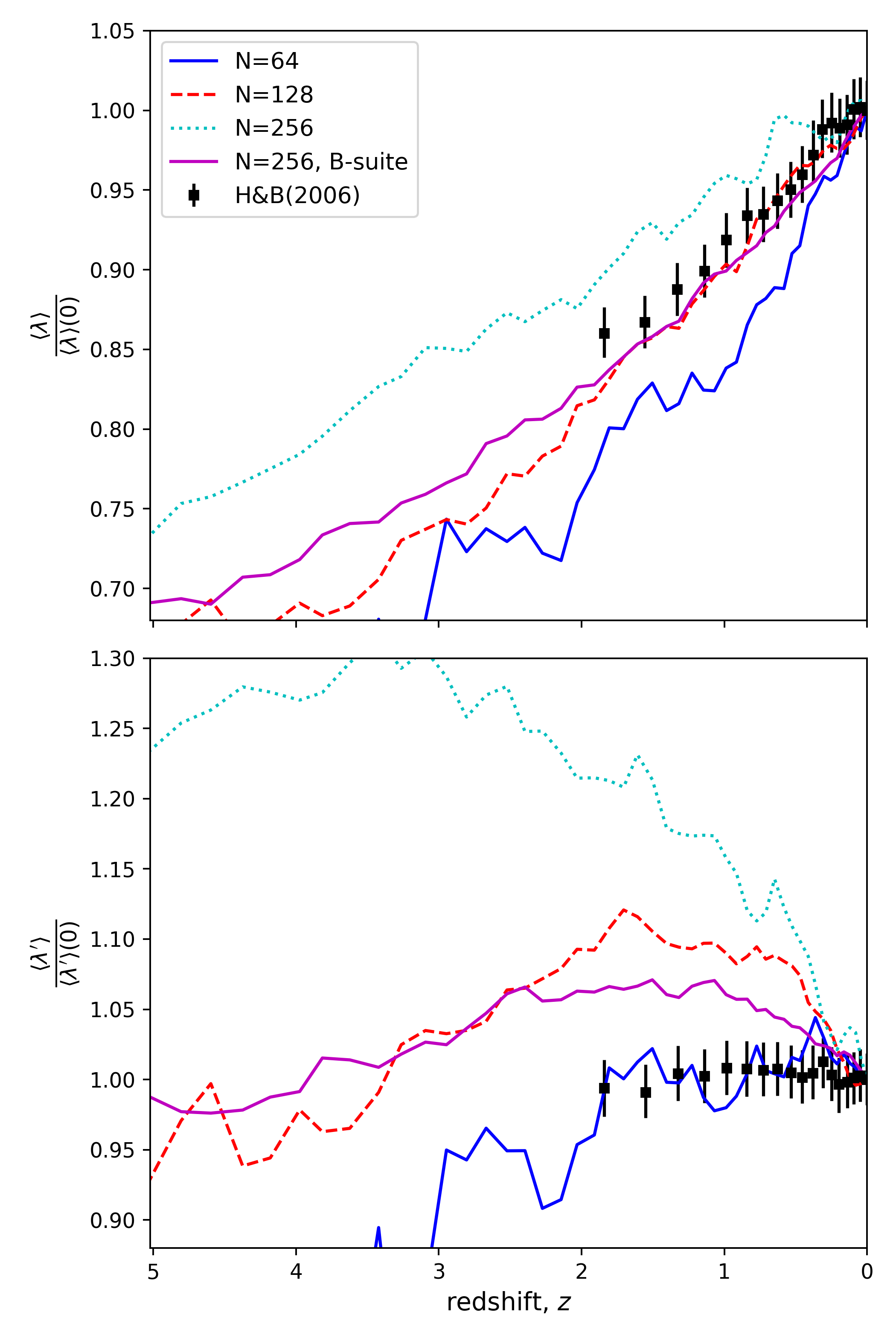}
\caption{Redshift evolution of the mean $\lambda$ (top panel) and $\lambda '$ (bottom panel), normalised by their values at redshift $z=0.0$, for our $B=20 \, Mpc \, h^{-1}$ $N=64^3$ 40-seed (blue solid curve), $128^3$ 10-seed (red dashed curve) and $256^3$ (cyan dotted curve) simulation sets, and our $N=256^3$ $B$-suite (purple solid curve). Using only host halos with $N_{p} \geq 500$. The \citet{HB06} data is also shown as black filled squares.}
\label{fig:N64vs128vs256}
}
\end{figure}

Figure~\ref{fig:N64vs128vs256} shows the redshift evolution of the mean spin parameters $\lambda$ (top panel) and $\lambda'$ (bottom panel), normalised by their respective $z=0.0$ values. The figure includes results from the $N=64^3$ ensemble (solid blue curve), the $N=128^3$ ensemble (dashed red curve), a single $N=256^3$ particle $B=20,\unit{Mpc,h^{-1}}$ simulation (cyan dotted curve), and the full $N=256^3$ particle $B$-suite (solid purple curve). For reference, the original \citet{HB06} data are shown as black filled squares.

Note that the $N=256^3$ $B$-suite results correspond to the $N_p^{\mathrm{min}}=500$ threshold introduced in Section~\ref{subsec:mean_spin_z} and illustrated in Figure~\ref{fig:meanlog_Ncuts_R200}. The agreement between our $N=64^3$ ensemble and the \citet{HB06} data is expected and reassuring.

We find that increasing resolution affects the behaviour of the spin parameters. The effect is modest for $\lambda$, and more pronounced for $\lambda'$, which we find to be more sensitive to both particle number $N$ and box size $B$. This is consistent with its dependence on halo boundary definitions and dynamical state. As shown by \citet{Davis10}, beyond $N=256^3$ particles, the spin probability distribution becomes largely insensitive to further increases in resolution, suggesting that our results are robust.

An aspect not widely explored in the literature \citep[but see][]{Power06} is the influence of simulation volume. As seen in Figure~\ref{fig:N64vs128vs256}, box size $B$ has a noticeable impact on the evolution of $\lambda'$ in particular. This warrants further investigation in future work. Overall, we successfully reproduce the behaviour reported by \citet{HB06} and demonstrate convergence toward our main results with increasing resolution and simulation volume.

\end{document}